\pdfoutput=1
\documentclass[aps,prd,twocolumn,superscriptaddress,tightenlines,nofootinbib]{revtex4}
\usepackage{graphicx}
\usepackage{epsfig}
\usepackage{bm}
\usepackage{latexsym,amssymb,amsmath,amsfonts,amssymb,txfonts,pxfonts,wasysym,float}
\usepackage{color}

\newcommand{\beq}[1]{\begin{equation}\label{#1}}
\newcommand{\eeq}{\end{equation}}
\newcommand{\bea}[1]{\begin{eqnarray} \label{#1}}
\newcommand{\eea}{\end{eqnarray}}
\newcommand{\ba}{\begin{array}}
\newcommand{\ea}{\end{array}}

\def\be{\begin{equation}}
\def\ee{\end{equation}}
\def\gs{\mathrel{
   \rlap{\raise 0.511ex \hbox{$>$}}{\lower 0.511ex \hbox{$\sim$}}}}
\def\ls{\mathrel{
   \rlap{\raise 0.511ex \hbox{$<$}}{\lower 0.511ex \hbox{$\sim$}}}}

\newcommand{\postscript}[2]{\setlength{\epsfxsize}{#2\hsize}
   \centerline{\epsfbox{#1}}}

\newcommand{\comment}[1]{}

\usepackage[usenames,dvipsnames]{xcolor}
\definecolor{orange}{cmyk}{0,0.5,1,0}
\definecolor{rossoCP3}{cmyk}{0,.88,.77,.40}
\definecolor{graa}{rgb}{0.8,0.8,0.8}
\definecolor{blaa}{rgb}{0.2,0.2,0.6}

\begin{document}

\title{\color{rossoCP3}{New test of Lorentz symmetry using
    ultrahigh-energy cosmic rays
}}

\author{Luis A. Anchordoqui}
\affiliation{Department of Physics \& Astronomy,  Lehman College, City University of
  New York, NY 10468, USA}
\affiliation{Department of Physics,
 Graduate Center, City University
  of New York,  NY 10016, USA}
\affiliation{Department of Astrophysics,
 American Museum of Natural History, NY
 10024, USA}

\author{Jorge F. Soriano}
\affiliation{Department of Physics \& Astronomy,  Lehman College, City University of
  New York, NY 10468, USA}
\affiliation{Department of Physics,
 Graduate Center, City University
  of New York,  NY 10016, USA}

\begin{abstract}
  \noindent We propose an innovative test of Lorentz symmetry by
  observing pairs of simultaneous parallel extensive air showers
  produced by the fragments of ultrahigh-energy cosmic ray nuclei
  which disintegrated in collisions with solar photons. We show that
  the search for a cross-correlation of showers in arrival time and
  direction becomes background free for an angular \mbox{scale $\alt
    3^\circ$} and a time window ${\cal O}(10~{\rm s})$. We also show
  that if the solar photo-disintegration probability of helium is
  ${\cal O} (10^{-5.5})$ then the hunt for spatiotemporal coincident
  showers could be within range of existing cosmic ray facilities, such
  as the Pierre Auger Observatory. We demonstrate that the actual
  observation of a few events can be used to constrain Lorentz
  violating dispersion relations of the nucleon.
\end{abstract}

\maketitle

Ever since Greisen, Zatsepin, and Kuzmin (GZK) pointed out that the
pervasive radiation fields make the universe opaque to the propagation
of ultrahigh-energy ($E \agt 10^9~{\rm GeV}$) cosmic rays
(UHECRs)~\cite{Greisen:1966jv,Zatsepin:1966jv}, it became evident that
the actual observation of the GZK effect would provide strong
constraints on Lorentz invariant breaking effects. This is because if
Lorentz invariance is broken in the form of non-standard dispersion
relations for various particles, then absorption and energy loss
processes for UHECR interactions would be modified; see
e.g.~\cite{Coleman:1998ti,Aloisio:2000cm,Jankiewicz:2003sm,Galaverni:2007tq,Galaverni:2008yj,Mattingly:2009jf,Scully:2010iv,Saveliev:2011vw,Stecker:2017gdy}. In
particular, the GZK interactions (photo-pion production and nucleus
photo-disintegration) are characterized by well defined energy
thresholds (near the excitation of the $\Delta^+ (1232)$ and the giant
dipole resonance, respectively), which can be predicted on the basis
of Lorentz invariance. Therefore, the experimental confirmation that
UHECR processes occur at the expected energy thresholds can be
considered as an indirect piece of evidence supporting Lorentz
symmetry under colossal boost transformations.

A suppression in the UHECR flux at $E \agt 10^{10.6}~{\rm GeV}$ has
been established beyond no doubt by the HiRes~\cite{Abbasi:2007sv},
Auger~\cite{Abraham:2008ru}, and Telescope Array
(TA)~\cite{AbuZayyad:2012ru} experiments.  By now (in Auger data) the
suppression has reached a statistical significance of more than
$20\sigma$~\cite{Abraham:2010mj}. This suppression is consistent with
the GZK prediction that interactions with universal photon fields
will rapidly degrade the energy of UHECRs. Intriguingly, however,
there are also indications that the source of the suppression may be
more complex than originally anticipated.

Observations of the rate of change with energy of the mean
depth-of-shower-maximum $X_{\rm max}$ seem to indicate that the cosmic
ray composition becomes lighter as energy increases toward $E \sim
10^{9.3}~{\rm GeV}$ from below~\cite{AbuZayyad:2000ay}, fueling a
widespread supposition that extragalactic cosmic rays are primarily
protons. However, Auger high-quality, high-statistics data, when
interpreted with the leading LHC-tuned shower models, exhibit a strong
likelihood for a composition that becomes gradually heavier with
increasing energy; namely, $1.5 \alt \langle \ln A \rangle \alt 3$,
for $10^{9.5} \alt E \alt
10^{10.6}$~\cite{Abraham:2010yv,Aab:2014kda,Aab:2014aea,Aab:2016htd}. Within
uncertainties, the data from TA are consistent with these
findings~\cite{Abbasi:2014sfa, Abbasi:2015xga}. For $E \agt
10^{10.6}~{\rm GeV}$, the indication of an anisotropy at an
intermediate angular scale of $13^\circ$ (significant at the
$4.0\sigma$ level~\cite{note})~\cite{Aab:2017njo} points to a similar
nuclear composition.  Note that for $E/Z = 10^{10}~{\rm GeV}$, typical
deflections of UHECRs crossing the Galaxy are about $10^\circ$, where
$Ze$ is the nucleus charge~\cite{Aab:2017tyv}.

For a uniform source
distribution, the simultaneous fit to the UHECR spectrum and
composition ($X_{\rm max}$ and its fluctuations) imposes severe
constraints on model parameters: {\it (i)} hard source spectra and
{\it (ii)}~a maximum acceleration energy $E_{\rm max} \alt 10^{9.7} \, Z~{\rm
  GeV}$~\cite{Aloisio:2013hya,Unger:2015laa,Aab:2016zth}. Hence, under the
assumption of a uniform source distribution,  the data seem to favor
the so-called ``{\sl disappointing} model''~\cite{Aloisio:2009sj} wherein it is
postulated that the ``end-of steam'' for cosmic accelerators is
coincidentally near the putative GZK cutoff, with the exact energy
cutoff determined by data. This interpretation encompasses a radically
different viewpoint in which the maximum energy of the most powerful
cosmic ray accelerators would be observed for the first time, and
therefore could call into question limits on the violation of Lorentz
invariance deduced using the observed suppression in the UHECR
spectrum~\cite{Scully:2008jp,Bi:2008yx,Aloisio:2014dua}.

Very recently, one of us put forward a multi-dimensional reconstruction of
the individual emission spectra (in energy, arrival direction, and
nuclear composition) to study the hypothesis that primaries are heavy
nuclei subject to GZK photo-disintegration, and to determine the
nature of the extragalactic sources~\cite{Anchordoqui:2017abg}. In
this paper we introduce an alternative approach to probe Lorentz
invariance using UHECRs. We propose to search for a cross-correlation
in arrival time and direction of the secondary nucleon (of energy
$E/A$) produced via photo-disintegration of an UHECR nucleus (of
energy $E$ and baryon number $A$) and the associated surviving
fragment (of baryon number $A-1$). Such a correlation study is
possible because: {\it (i)}~the Lorentz factor (which is equivalent to
energy per nucleon) is conserved for photo-disintegration and {\it
  (ii)} the trajectory of cosmic rays within a magnetic field is only
rigidity-dependent; the relevant quantity for the separation among
fragments (hereafter identified with subindices 1 and 2)  is  $|Z_1/A_1 - Z_2/A_2|$.

A simple dimensional argument constrains the distance to the
photo-disintegration site.  Assuming the energy difference between
nucleons inside the nucleus is given by the binding energy $E_0 \sim
{\rm MeV}$, the difference in velocity of the secondary products is
\begin{equation}
\delta v = \sqrt{ 2 E_0/M} \sim \sqrt{10^{-3}/A} \,,
\end{equation}
where $M \simeq A~{\rm GeV}$ is the mass of the parent nucleus. The
difference in the time of flight of the secondary products is then
\begin{equation}
\delta t \sim \delta L = \frac{(L/{\rm Mpc})}{\gamma} \ \delta v \times 10^{24}~{\rm
  cm} \,,
\end{equation}
where $L$ is the distance to the photo-disintegration site  and $\gamma$
($= E/M$ at Earth) contracts this length. For a simultaneous
observation  of the two secondaries at Earth, we demand 
$\delta L \alt 2 \, R_\oplus \ (\sim 10^9~{\rm cm})$, which yields
\begin{equation}
\gamma \sim \frac{10^{14}}{ \sqrt{10 \ A}} \ \ (L/{\rm Mpc}) \, .
\label{gamma}
\end{equation}
For the particular range $10^9 \alt \gamma \alt 10^{10}$, which spans
the UHECR spectrum, (\ref{gamma}) constrains the photo-disintegration
site to a \mbox{distance $\alt {\rm kpc}$.}  It has long been known
that UHECR nuclei scattering off the universal radiation fields have a mean
free path $\gg {\rm kpc}$~\cite{Puget:1976nz}. Moreover, we know the
devil is in the detail and so the number of GZK interactions which
would lead to a simultaneous observation of their secondary products
on Earth is essentially negligible.

Of particular interest here, UHECR nuclei {\it en route} to Earth also
interact with the solar radiation field and
photo-disintegrate~\cite{Zatsepin,Gerasimova}. The nuclear
photo-disintegration process has two characteristic regimes. There is
the domain of the giant dipole resonance (GDR), where a collective
nuclear mode is excited with the subsequent emission of one (or
possible two nucleons), and the high energy plateau, where the excited
nucleus decays dominantly by two nucleon and multi-nucleon
emission. The energy range of the GDR in the nucleus rest frame spans
$10 \alt \varepsilon'/{\rm MeV} \alt 30$, and the plateau extends up
to the photo-pion production threshold (i.e., photon energy
$\varepsilon' \sim 150~{\rm MeV}$). 

The background radiation field can be described by a Planckian
spectrum, with a  temperature of the solar surface $T_s \simeq
0.5~{\rm eV}$, normalized to reproduce the solar luminosity, $L_\odot
= 4 \pi r^2 c \int d\varepsilon \ \varepsilon  \ dn/d\varepsilon$, yielding
\begin{equation}
\frac{dn}{d\varepsilon} = 7.2 \times 10^7
\frac{\varepsilon^2}{\exp(\varepsilon/T_s) -1 } \left( \frac{r}{{\rm
      AU}} \right)^{-2}~{\rm (eV \, cm)^{-3}} \,,
 \end{equation}
where $r$ is the spherical radial coordinate centered at the Sun.
In the rest frame of the nucleus, the energy $\varepsilon$ of the solar
photons (in the rest frame of the Sun) is highly blue-shifted to
\begin{equation}
\varepsilon' = \varepsilon \gamma (1 + \beta \ \cos \alpha) \sim 2 \gamma \
\varepsilon \ c_{\alpha/2}^2 \,,
\end{equation}
where $\beta = \sqrt{1 - 1/\gamma^2} \sim 1$ and $c_{\alpha/2} = \cos(
\alpha(\ell)/2)$, and where $\alpha(\ell)$ is the angle between the
momenta of photon and nucleus in the Sun's reference frame, with
$\ell$ the coordinate along the path of the nucleus; i.e., $\cos
\alpha = \hat \ell \cdot \hat r$.

The GDR cross section in the narrow width approximation is
\begin{equation}
\sigma (\varepsilon) = \frac{\pi}{2} \ \sigma_0  \ \Gamma \ \delta (2
\gamma \ \varepsilon \ c_{\alpha/2}^2  -
\varepsilon_0) \,,
\end{equation}
where $\Gamma$ and $\sigma_0$ are the GDR width and cross section at
maximum; the factor of 1/2 is introduced to match the integral
(i.e. total cross section) of the Breit-Wigner and the delta
function~\cite{Anchordoqui:2006pd}. Fitted numerical formulas are
$\sigma_0 = 1.45 A~{\rm mb}$, $\Gamma = 8~{\rm
  MeV}$, and $\varepsilon_0 = 42.65A^{-0.21}~{\rm MeV}$ for $A > 4$ and
$\varepsilon_0 = 0.925A^{2.433}~{\rm MeV}$ for $A\leq
4$~\cite{Karakula:1993he}. In the high energy regime the cross section
is well approximated by $\sigma
(\varepsilon) \approx A/8~{\rm mb}$.

All in all, the probability that a
nucleus photo-disintegrates on the solar radiation along its path towards
the Earth is found to be
\begin{equation}
\eta_A = 1 - \exp \left[ - \int_0^\infty d \ell \ \frac{1}{\lambda (\ell)}
\right] \,,
\label{ocho}
\end{equation}
where 
\begin{equation}
\frac{1}{\lambda (\ell)} = \int_0^\infty  \sigma
(\varepsilon) \ \ \frac{dn}{d \varepsilon}  \ \ 2  \ c^2_{\alpha/2}  \ d\varepsilon 
\end{equation}
is the inverse photo-disintegration
mean-free-path~\cite{MedinaTanco:1998ac}.  Integration of (\ref{ocho})
yields: $10^{-5} \alt \eta_A \alt 10^{-4}$ for iron, and $10^{-6}
\alt \eta_A \alt 10^{-5}$ for helium and oxygen. These values of $\eta_A$
are in agreement with the estimates 
in~\cite{MedinaTanco:1998ac,Epele:1998mv,Lafebre:2008rz}.

Since the secondary fragments have slightly different rigidities the
deflection in the interplanetary magnetic field will result in two
separate extensive air showers, arriving essentially at the same time
and from the same direction in the
sky~\cite{MedinaTanco:1998ac,Epele:1998mv}.  More specifically, the
average separation of the shower on Earth can be parametrized by~\cite{Lafebre:2008rz}
\begin{equation}
\langle \delta L \rangle_A = 4 A  \left|\frac{Z_1}{A_1} - \frac{Z_2}{A_2} \right|
  \left(\frac{E}{10^{10}~{\rm GeV}} \right)^{-1}~{\rm km} \, ,
\label{delta}
\end{equation} where $E$ is the energy of the parent nucleus.
The average separation of showers
as estimated in~\cite{Epele:1998mv} is somewhat smaller.  For a given 
experiment, each nuclear species has a
critical energy above which $\langle \delta L \rangle_A$ would be
comparable to the size of the instrumented area. As benchmark we
consider a $3,000~{\rm km}^2$ array of detectors, with interspacing of
about 1.5~km. For $^4$He,
(\ref{delta}) yields $\langle \delta L \rangle_{\rm He} \sim 50~{\rm
  km}$ at $E \sim 10^{9.3}~{\rm GeV}$. However, for $^{56}$Fe, at the
same energy (\ref{delta}) leads to $\langle \delta L \rangle_{\rm Fe}
\sim 260~{\rm km}$, and so the
separation distance between the showers would be out of detection range.

Because the intensity of cosmic rays is steeply falling with energy,
contributions from the counting rate at the critical energy dominate
the integrated event rate. Existing estimates of the event rate at
UHECR
facilities~\cite{MedinaTanco:1998ac,Epele:1998mv,Lafebre:2008rz,vanEijden:2016yqq}
are subject to large uncertainties, mainly because  
$\eta_A$ and $\langle \delta L \rangle_A$ depend strongly on $A$ and
the nuclear composition of UHECRs is poorly known.

Herein, we assume a nuclear composition dominated by helium at $E \agt
10^{9.3}~{\rm GeV}$ that becomes gradually heavier with increasing
energy; see e.g. Fig. 4 of \cite{Unger:2015laa}. We further assume that the photo-disintegration probability of
helium on the solar photons is $\eta_{\rm He} \sim 10^{-5.5}$. These
two assumptions together lead to an expected integrated flux of
\begin{equation}
\frac{d F}{dt \ d\Omega \  dA} (E> 10^{9.3}~{\rm GeV}) \sim 3 \times
10^{-5}~{\rm km}^{-2} \, {\rm sr}^{-1} \, {\rm yr}^{-1} \,,
\label{flux}
\end{equation}
where $E$ denotes the energy of the parent nucleus. This flux is in
agreement with the one shown in Fig.~3
of~\cite{Epele:1998mv}. Moreover, as
exhibited in Fig. 2 of~\cite{Epele:1998mv}, for $\eta_{\rm He} \sim
10^{-5.5}$ and $E \agt 10^9~{\rm GeV}$, we have $20 \alt \langle \delta
L \rangle_{\rm He}/{\rm km} \alt 50$. The flux derived herein,
using a helium saturated spectrum above $10^{9.3}~{\rm GeV}$, is
larger than the intensity derived in~\cite{Lafebre:2008rz} using the
spectrum of~\cite{Hoerandel:2002yg}. Whichever flux calculation one
may find more convincing, it seems most conservative at this point to
depend on experiment (if possible) to resolve the issue.

The $3,000~{\rm km}^2$ surface detector array of the Pierre Auger
Observatory is fully efficient at $E \agt 10^{9.5}~{\rm GeV}$~\cite{Abraham:2010zz}. From
January 2004 until December 2016 this facility has accumulated an
exposure~\cite{Aab:2017njo}
\begin{equation}
{\cal E}  (E > 10^{9.5}~{\rm GeV})= 6.7 \times 10^4~{\rm km^2 \, sr \,
  yr} \, . 
\end{equation}
At lower energies, the trigger efficiency of the surface detector
array decreases smoothly and becomes roughly 30\% at $10^{8.7}~{\rm
  GeV}$~\cite{Abraham:2010zz}. To get a rough estimate of the exposure
available to probe spatiotemporal correlations of air showers in an
experiment like Auger we scale down ${\cal
  E} (E > 10^{9.5}~{\rm GeV})$ by a factor of 0.3. This leads to
\begin{equation}
{\cal E} (10^{8.7} < E/{\rm GeV} < 10^{9.3}) \agt 2 \times 10^4~{\rm
  km^2 \, sr \, yr} \, .
\label{exposure}
\end{equation}
For $^4$He, $\Delta E = E_2 - E_1 \sim 3\gamma~{\rm GeV}$.
For $E> 10^{9.3}~{\rm GeV}$, (\ref{flux}) and (\ref{exposure}) lead to an expected integrated rate
which is consistent with 1 event.

\begin{figure*}[tbp]
\begin{minipage}[t]{0.33\textwidth}
\postscript{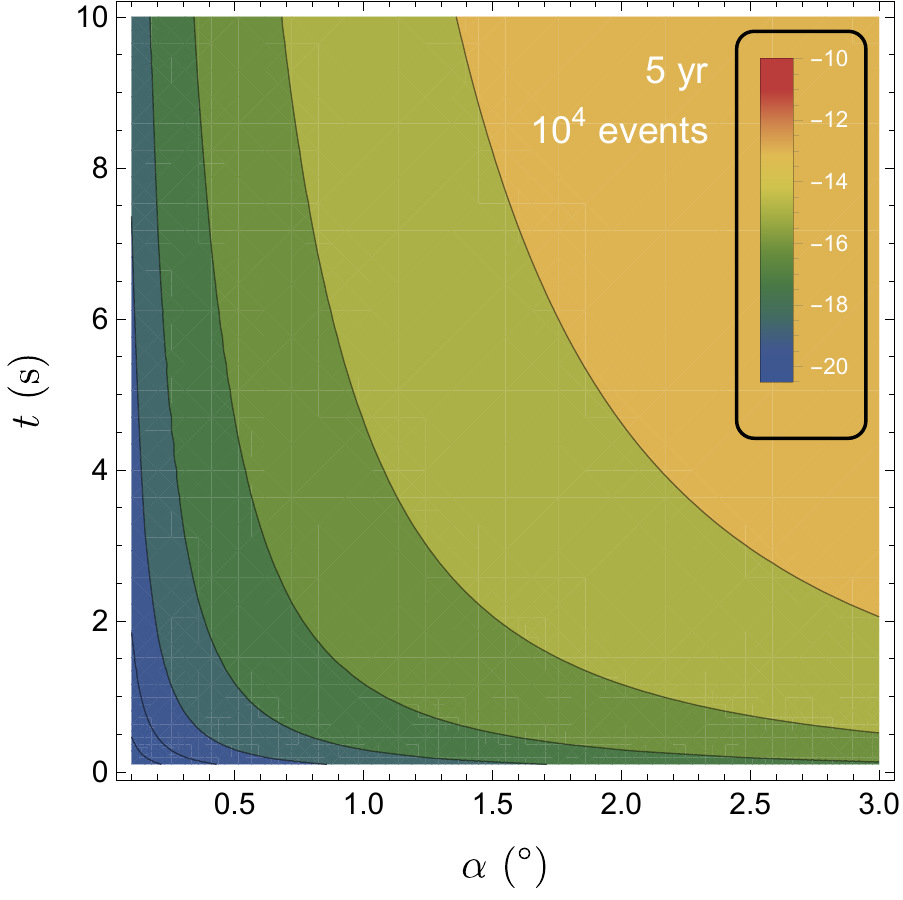}{0.99}
\end{minipage}
\begin{minipage}[t]{0.33\textwidth}
\postscript{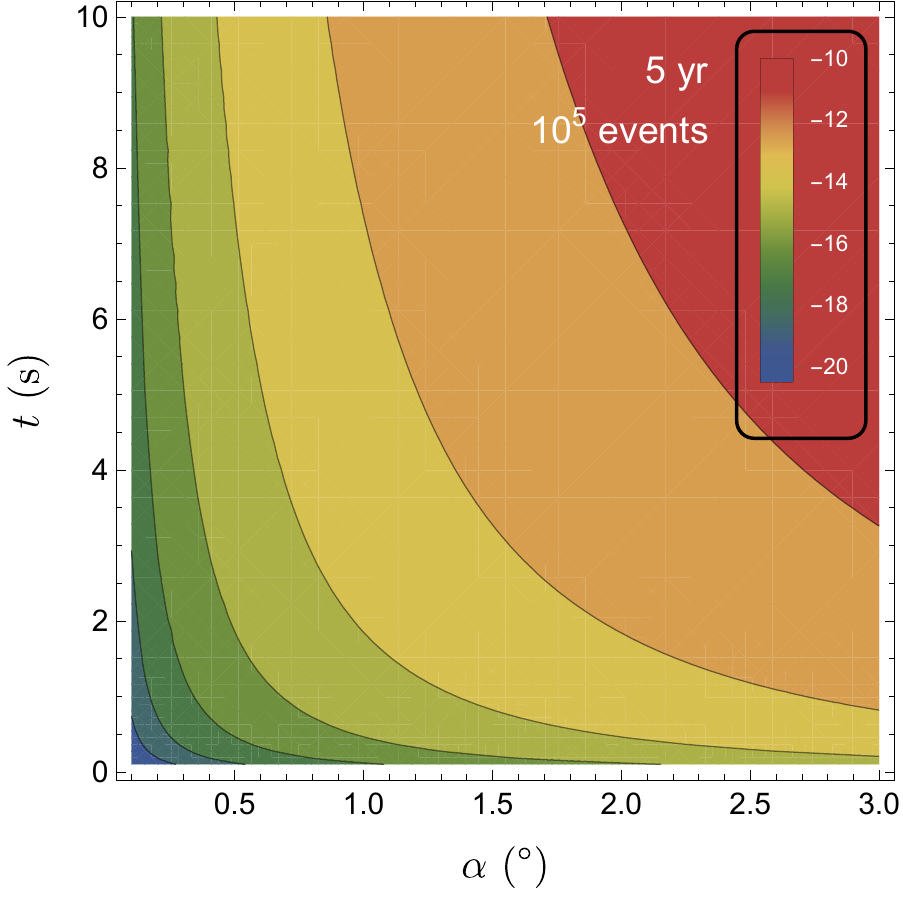}{0.99}
\end{minipage}
\begin{minipage}[t]{0.33\textwidth}
\postscript{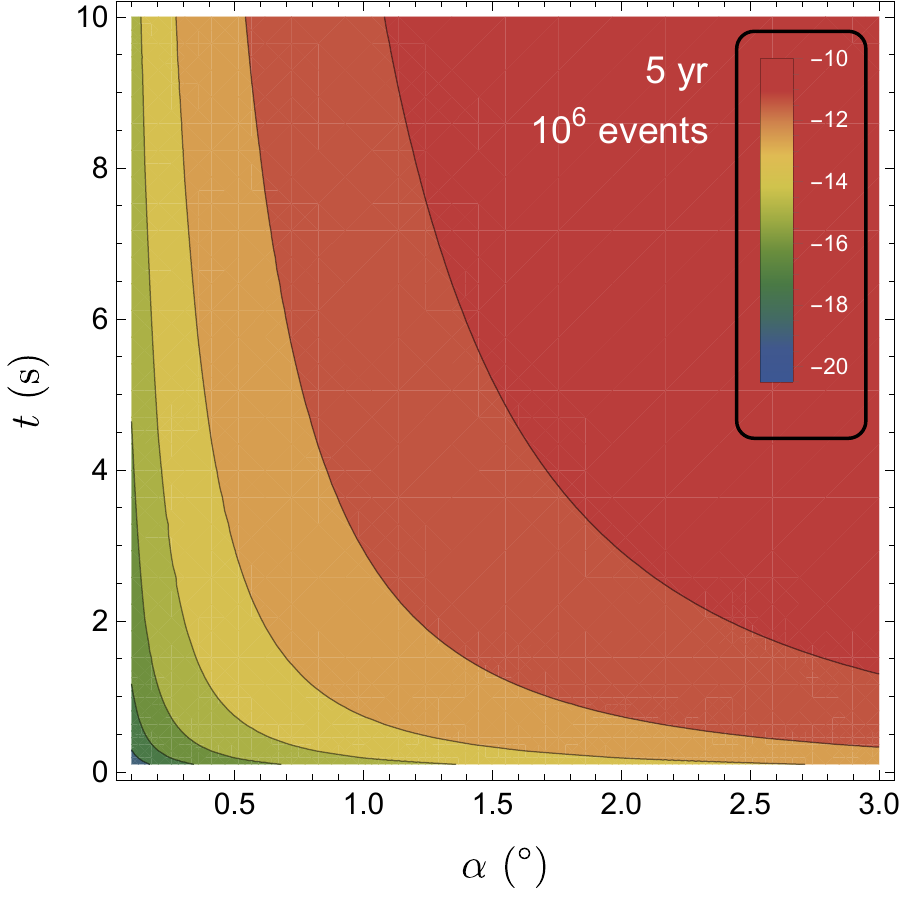}{0.99}
\end{minipage}
\begin{minipage}[t]{0.33\textwidth}
\postscript{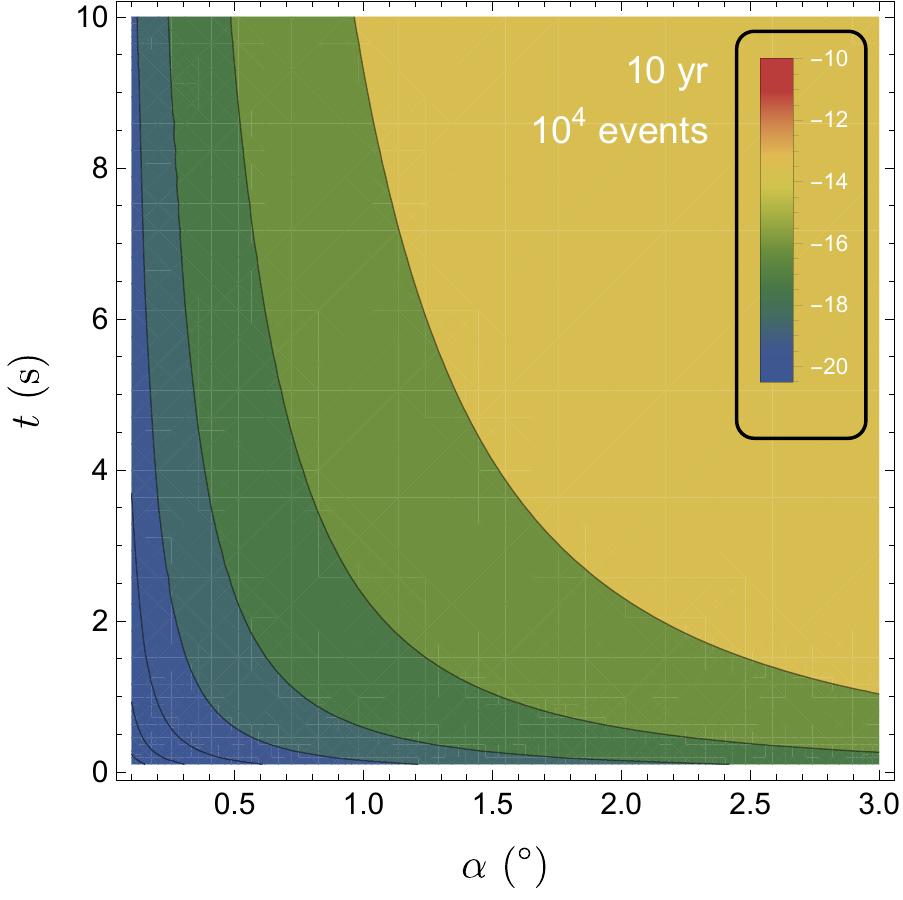}{0.99}
\end{minipage}
\begin{minipage}[t]{0.33\textwidth}
\postscript{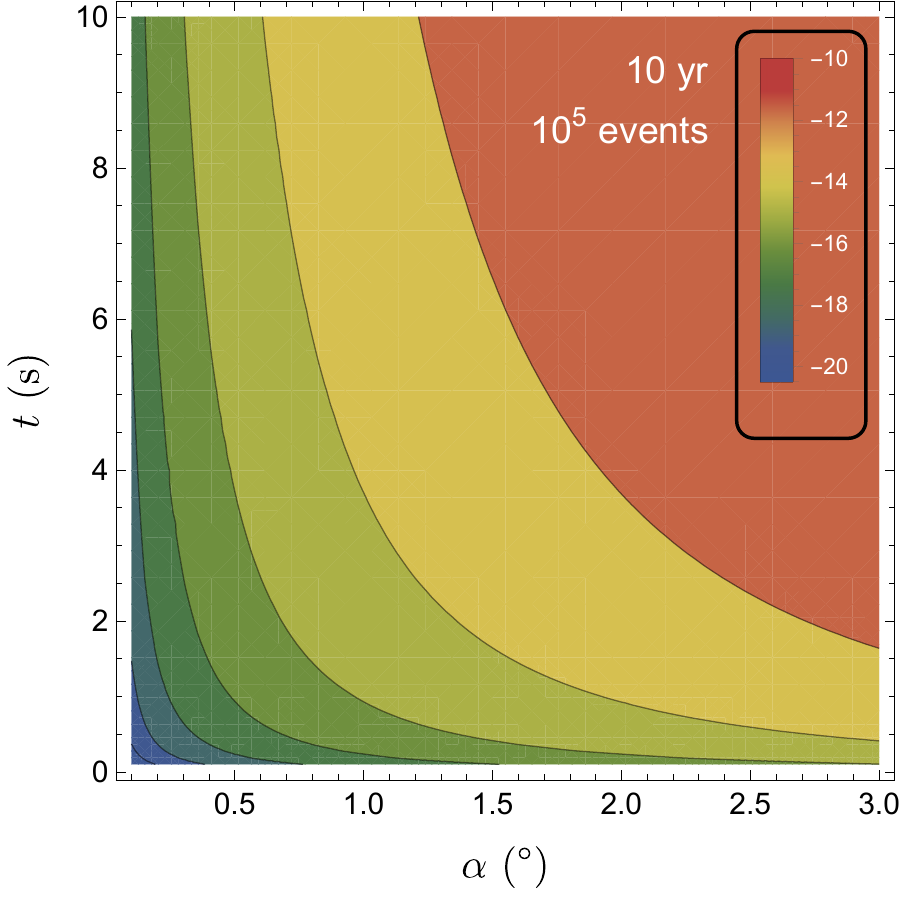}{0.99}
\end{minipage}
\begin{minipage}[t]{0.33\textwidth}
\postscript{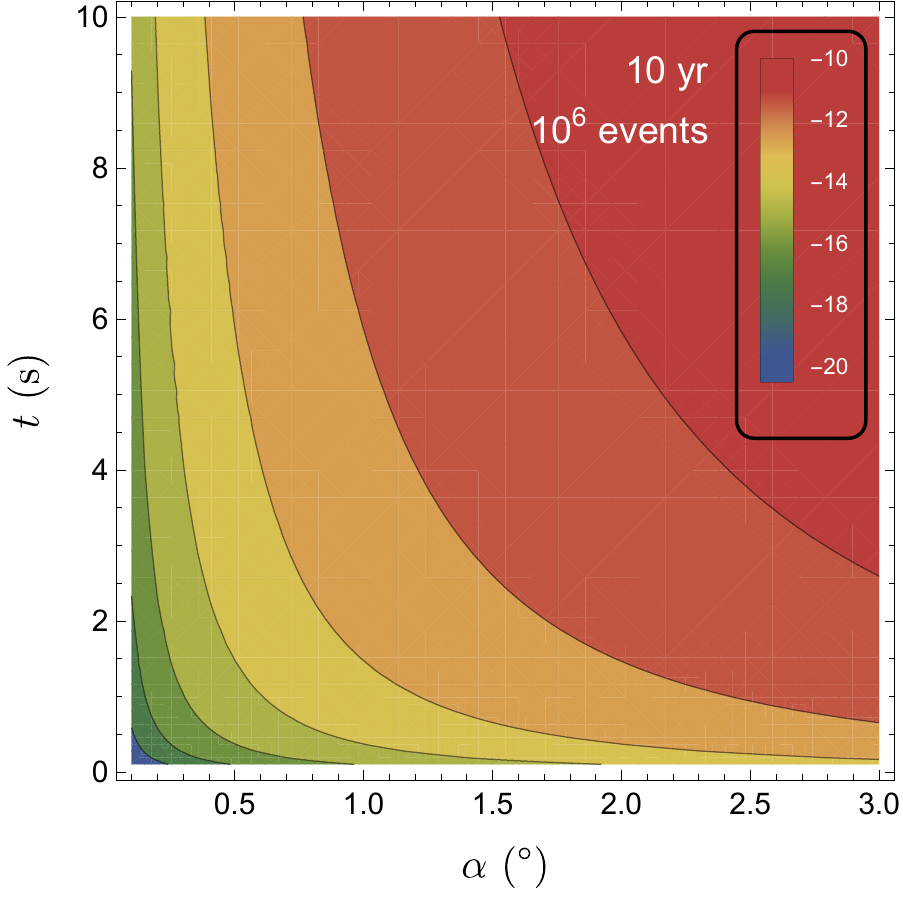}{0.99}
\end{minipage}
\begin{minipage}[t]{0.33\textwidth}
\postscript{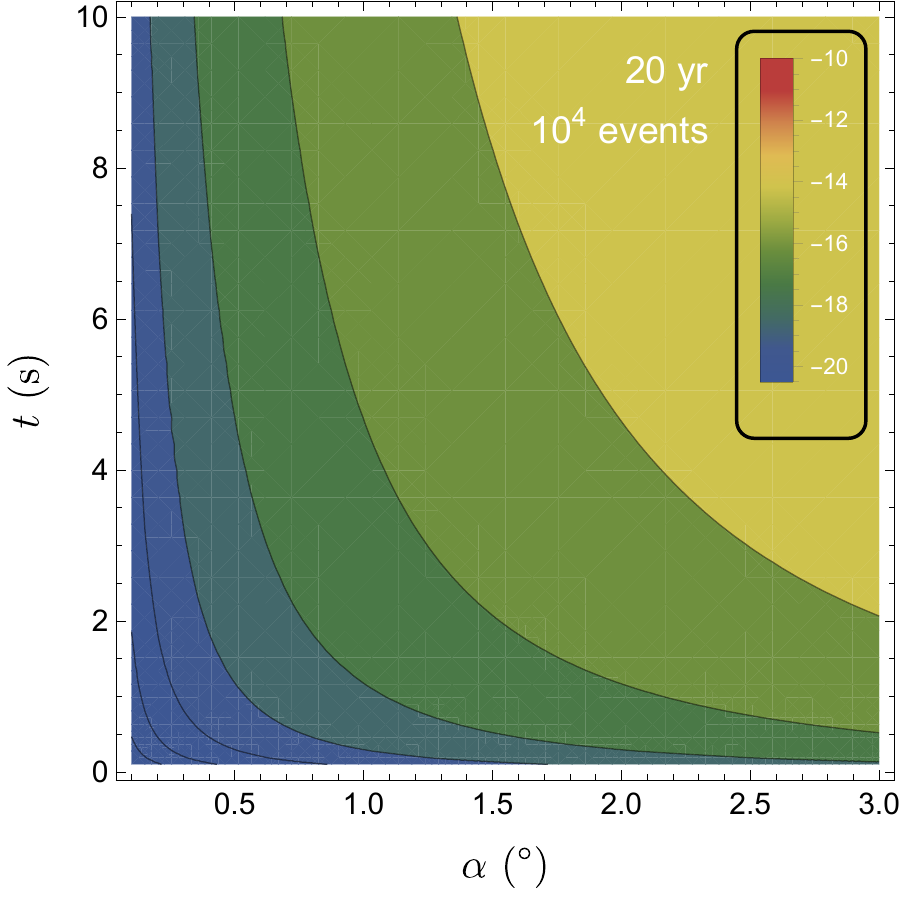}{0.99}
\end{minipage}
\begin{minipage}[t]{0.33\textwidth}
\postscript{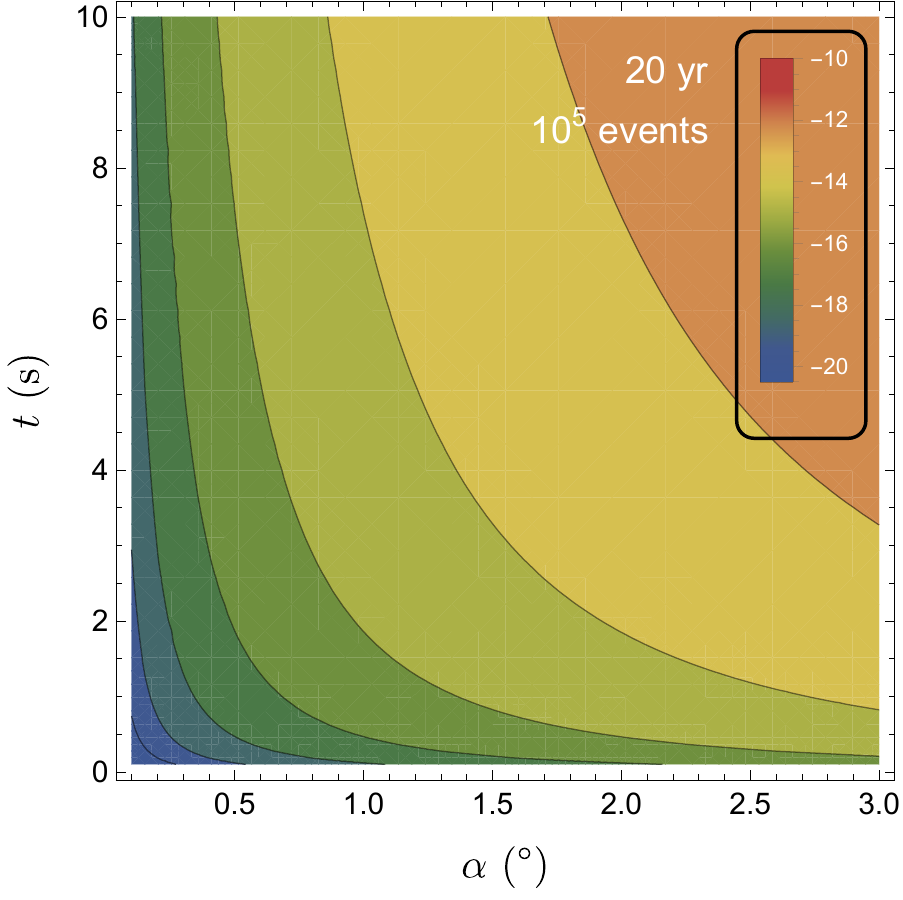}{0.99}
\end{minipage}
\begin{minipage}[t]{0.33\textwidth}
\postscript{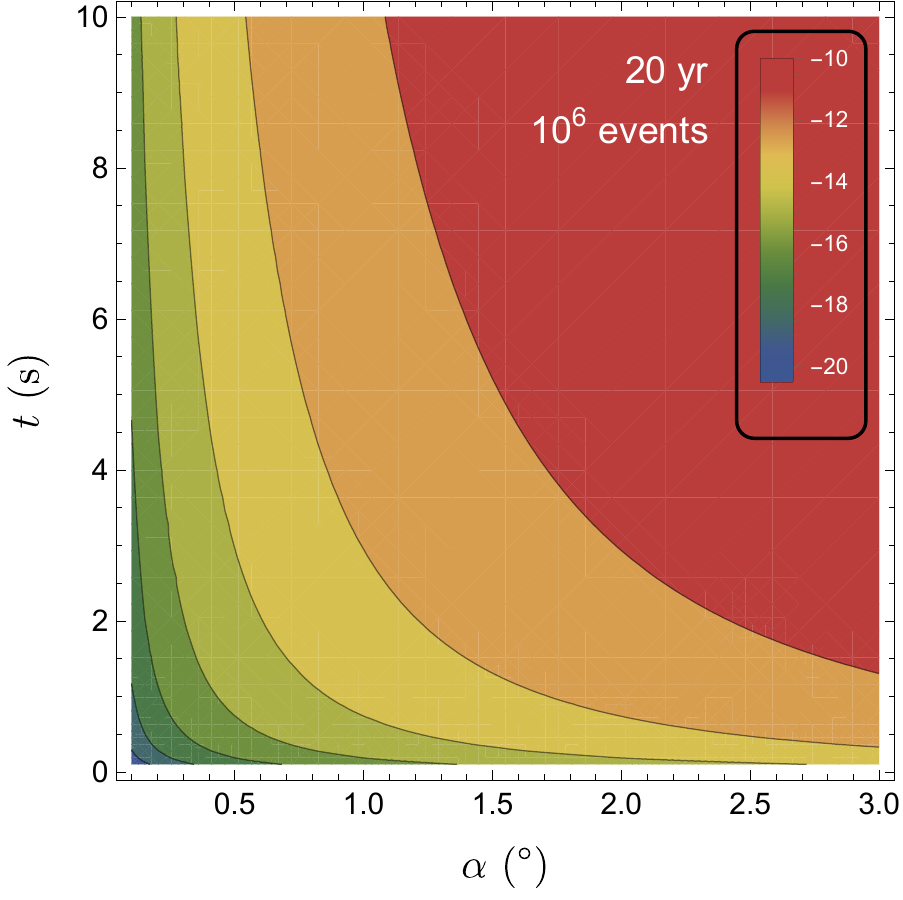}{0.99}
\end{minipage}
\caption{$\log_{10} p_2(\alpha,t)$ for different total number of
  events (from left ro right $N = 10^4, 10^5, 10^6$), and different lifetimes
  of the experiment (from top to bottom $T/{\rm yr} = 5, 10, 20$). $T=10\,\mathrm{yr}$ and $T=20\,\mathrm{yr}$ are scales compatible with  Auger, while $T=5\,\mathrm{yr}$ is an approximation for the life span of the prospective experiment POEMMA~\cite{Olinto:2017xbi}.
\label{fig:1}}
\end{figure*}

It is clear that for a signal  ${\cal O} (1)$ event 
we must learn how to properly conduct background rejection to
ascertain whether the observation of a few events is due to {\it physics} or
{\it statistics}. Moreover, to calculate a meaningful statistical
significance in the shower cross-correlation analysis, it is important
to define the search procedure {\it a priori} in order to ensure it is
not inadvertently devised especially to suit the particular data set
after having studied it. With the aim of avoiding accidental bias on
the number of trials performed in selecting the cuts,  we now conduct a
phenomenological analysis of the potential background to define the
angular and temporal cuts.

We start by selecting a reference direction on the sky $\mathfrak
d_0$. We define $\theta$ as the angle between $\mathfrak d_0$ and
other direction on the sky $\mathfrak d$.  We define $\phi$ as the
angular distance between a reference axis, placed on the normal plane
to the vector pointing in the direction $\mathfrak d_0$, and the
projection on that plane of a vector pointing towards $\mathfrak
d$. With this construction, $\theta\in[0,\pi]$ and $\phi\in[0,2\pi]$.

The expected fraction of events that will be contained in a cap of the
sphere of within an angle $\alpha$ to the direction $\mathfrak d_0$, for
all $\phi$, and in a time interval $t$ is 
\begin{eqnarray} 
  f(\alpha,t) & = &\frac{t}{T} \int_0^{2\pi}  d\phi \int_0^\alpha
   d \theta \frac{1}{4\pi}\sin\theta \nonumber \\
  & = & \frac{1}{2} \ \frac{t}{T} \ (1-\cos\alpha) \,,
\end{eqnarray} 
where $T$ is the time span for the experiment. In a sample of $N$
events, we expect $\mu(\alpha,t)=N f(\alpha,t)$ events in the
angle-time window defined above. The actual number of events in that window
 will be distributed following a Poisson distribution of mean
$\mu(\alpha,t)$. The probability of observing $k$ events in an
angle-time window is then
\begin{equation}
  p_k(\alpha,t)=\frac{e^{-\mu(\alpha,t)}\mu(\alpha,t)^k}{k!}.
\end{equation}
In Fig.~\ref{fig:1} we show the function $\log_{10}p_2(\alpha,t)$, for
\mbox{$\alpha\in[0^\circ,3^\circ]$} and $t\in[0\,\mathrm s,10\,\mathrm s]$,
which gives the probability of measuring 2 events in an angle-time
window specified by the pair $(\alpha,t)$. Since \begin{equation}
  \frac{\sum_{k=3}^\infty
    p_k(\alpha,t)}{p_2(\alpha,t)}\lesssim10^{-6}\end{equation} in our
$(\alpha,t)$ range of interest, $p_2(\alpha,t)$ practically accounts
for the probability of having not only two, but any amount of events above
one.

\begin{figure}[tbp]
    \postscript{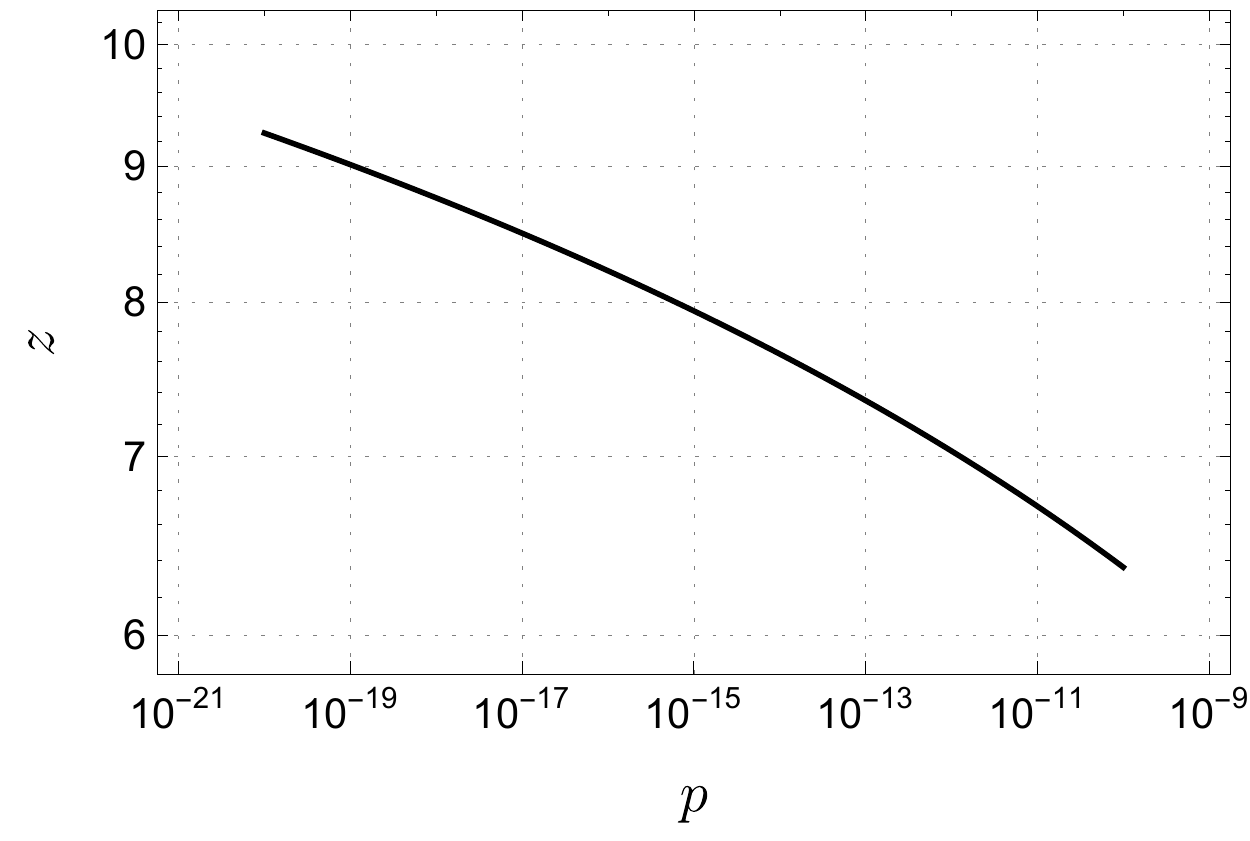}{0.99}
\caption{Relation between the $p$-value and $z$, the number of
  standard deviations away from the mean, for a normal distribution. 
\label{fig:2}}
\end{figure}

The quantity $p_2(\alpha,t)$ is then the $p$-value for observing a
coincidence of two detections in a background only hypothesis. To
quantify this in a more comprehensible way, one can use the usual
relation between $p$-values and $\sigma$ levels following a normal
distribution \begin{equation}
  p=\frac12\left[1-\mathrm{erf}\left(\frac{z}{\sqrt2}\right)\right],
\end{equation}
being $z$ the number of standard deviations from the mean. The
relation between $p$ and $z$ is shown in Fig.~\ref{fig:2}. One can
check by inspection that for the whole range of $\alpha$ and $t$, the
$p$-value for observing two or more events together in a
\textit{small} angle-time window is a more than $5$-sigma effect
against the background. Hence, the actual observation of a few pairs of
cross-correlated events would become the smoking gun to set model
independent constraints on Lorentz invariance violation.

Strictly speaking,  a nucleus with baryon number $A$ and charge $Ze$
would have a non-standard dispersion relation of the form
\begin{equation}
E^2_{A,Z} = p^2_{A,Z} + M_{A,Z}^2 + \zeta_{A,Z} \ 
\frac{p_{A,Z}^{n+2}}{M_{\rm Pl}^n} \,,
\label{eq:dispersion}
\end{equation}
where $E_{A,Z}$ is the nucleus energy, $p_{A,Z}$ is the absolute value
of its 3-momentum, and $M_{A,Z}$ its mass. Here, $M_{\rm Pl} \approx
10^{19}~{\rm GeV}$ is the Planck mass and $\zeta_{A,Z}$ are
Lorentz-violating parameters of the nucleus. In the rest frame of the Sun we assume that only
baryons have non-standard dispersion relations (note that the solar
photon fields have too low energy for Lorentz invariant breaking
effects to be relevant in their dispersion relations) and so one can easily
obtain a threshold relation, which constrains
the $\zeta_{A,Z}$ coefficients when confronted with data. Actually, since we expect nuclear physics to
have negligible Lorentz breaking effects we can write $\zeta_{A,Z} =
\zeta/A^2$, where $\zeta$ regulates deviations from Lorentz symmetry
of the nucleon. The baryon number $A$ of the original disintegrated
nucleus can simply be determined by estimating the energies of the
primaries of the two air showers, $A = 1+ E_2/E_1$, where $E_1$ is the
energy of the less energetic shower. With the $<20\%$ energy resolution achieved by the Pierre Auger Observatory~\cite{ThePierreAuger:2013eja}, the estimation of $A$ is obtained with a resolution $\sigma(A)/A<0.2\sqrt 2(1-1/A)$, which is around $20\%$ for a Helium primary, or $\sigma(A=4)\sim0.85$, allowing its differentiation from other primaries with $A$ around $4$. This provides an univocal (model
independent) determination of the nuclear composition and thereupon
bounds the threshold energy interval to be compatible with
experimental results on photo-nuclear
interactions~\cite{Gorbunov,Gorbunov:1968quk,Balestra:1977xn,McBroom:1982zz,Calarco:1983zz,Bernabei:1988rq,Feldman:1990zz,Shima:2005ix,Raut:2012zz,Tornow:2012zz}.

For $^4$He, the photo-excitation cross section of the GDR has a
threshold $\varepsilon'_{\rm th} \approx
20\,\mathrm{MeV}$~\cite{Stecker:1998ib}. The GDR decays by the
statistical emission of a single nucleon, leaving an excited daughter
nucleus $(A-1)^*$. The probability for emission of two (or more)
nucleons is smaller by an order of magnitude. The excited daughter
nuclei typically de-excite by emitting one or more photons of energies
$1 \alt \epsilon' /{\rm MeV} \alt 5$, in the nuclear rest
frame~\cite{Anchordoqui:2006pd}.  For simplicity, herein we neglect
the de-excitation process and consider the photo-disintegration
reaction with two incoming particles (nucleus + photon) and two
outgoing particles (nucleus + nucleon).  Though we are primarily
interested in helium photo-disintegration, the ensuing discussion will
be framed in a general context. The energy-momentum 4-vectors for the
four particles in the rest frame of the Sun are: $(E,\mathbf p)$, for
the incoming nucleus; $(\varepsilon,\mathbf k)$, for the photon;
$(E_1,\mathbf p_1)$, for the nucleon; and $(E_2,\mathbf p_2)$, for the
outgoing nucleus. The relation describing the conservation of energy
and momentum is given by
\begin{equation}
  (E+\varepsilon)^2-(\mathbf p+\mathbf k)^2=(E_1+E_2)^2-(\mathbf
  p_1+\mathbf p_2)^2 \, .
\label{eq:cons}
\end{equation} 
We are interested in studying
the energy thresholds for which the relation (\ref{eq:cons}) holds.

According to the threshold theorem, \textit{at an upper or lower
  threshold the incoming particle momenta are always anti-parallel and
  the final particle momenta are
  parallel}~\cite{Mattingly:2002ba}. This applies for dispersion
relations $E(\mathbf p)$ depending on $p\equiv|\mathbf p|$, and being
a monotonically increasing function of that variable, when energy and 
momentum are conserved additive quantities. Then, to obtain the
threshold conditions, one can make use of $\mathbf p\cdot\mathbf
k=-p\,k$ and $\mathbf p_1\cdot\mathbf p_2=p_1p_2$. Since we neglect
Lorentz invariant breaking effects on the solar photon fields we take
$\varepsilon=k$. In threshold conditions the reaction is
collinear and so $p-k=p_1+p_2$. Since $k$ is much smaller than the other
momenta, we have $p\approx p_1+p_2$. Following~\cite{Saveliev:2011vw},
we define
$p_2=\varkappa p$ and $p_1=(1-\varkappa)p$, with $0 < \varkappa < 1$.
Now, neglecting the mass difference between the proton and
the neutron ($M_{A,Z} =A\,m_p$, where $m_p$ is the proton mass), the
energy conservation relation is found to be
\begin{widetext}
\begin{equation} 
\xi_A(1)+\frac{2 \varepsilon}{p} \left[1+\sqrt{1+\xi_A(1)}
\right]  = \varkappa^2\xi_{A-1}(\varkappa)+(1-\varkappa)^2\xi_1
  (1-\varkappa)
+2\varkappa (1-\varkappa) 
\left[\sqrt{1+\xi_{A-1}(\varkappa)} \right. \left. \sqrt{1+\xi_{1}(1-\varkappa)}-1 \right] \,,
\end{equation}
\end{widetext} 
where
\begin{equation}
  \xi_A(\varkappa)=\left(\frac{A\,m_p}{\varkappa\,p}\right)^2+\frac{\zeta}{A^2}\left(\frac{\varkappa\,p}{M_{\rm
        Pl}}\right)^n \, .\end{equation} 
Note that for $M_{A,Z} \ll p_{A,Z} \equiv p \ll M_{\rm Pl}$,  
$\xi_A(\varkappa)\ll1$. Expanding the square roots to first order in the $\xi$ functions, 
\begin{equation}\left(1+\frac{\varepsilon}{p}\right) \xi_A(1)+ \frac{4
   \varepsilon}{p}=\varkappa\xi_{A-1}(\varkappa)+(1-\varkappa)\xi_1(1-\varkappa).
\label{ventiuno}
\end{equation}
Since all the $\xi$-functions are of the same order and
$\varepsilon\ll p$, the term $\varepsilon\xi_A(1)/p$ is negligible in
comparison to the rest of the terms, and so (\ref{ventiuno}) becomes
\begin{equation}
  \xi_A(1)+4\frac{\varepsilon}{p}=\varkappa\xi_{A-1}(\varkappa)+(1-\varkappa)\xi_1(1-\varkappa)
  \, .
\label{ventidos}
\end{equation}
After some algebra, (\ref{ventidos}) can be rewritten as 
\begin{equation}
\zeta \ g(\varkappa)\left(\frac{p}{m_p}\right)^2\left(\frac{p}{M_{\rm
      Pl}}\right)^n+\frac{4 \varepsilon
  p}{m_p^2}-\frac{[1-(1-\varkappa)A]^2}{\varkappa(1-\varkappa)}=0 \,,
\end{equation}
where
\begin{equation}
g (\varkappa) =
\frac{1}{A^2}-\frac{\varkappa^{n+1}}{(A-1)^2}-(1-\varkappa)^{n+1} 
\, .
\end{equation}
\begin{figure*}[tbp]
\begin{minipage}[t]{0.49\textwidth}
\postscript{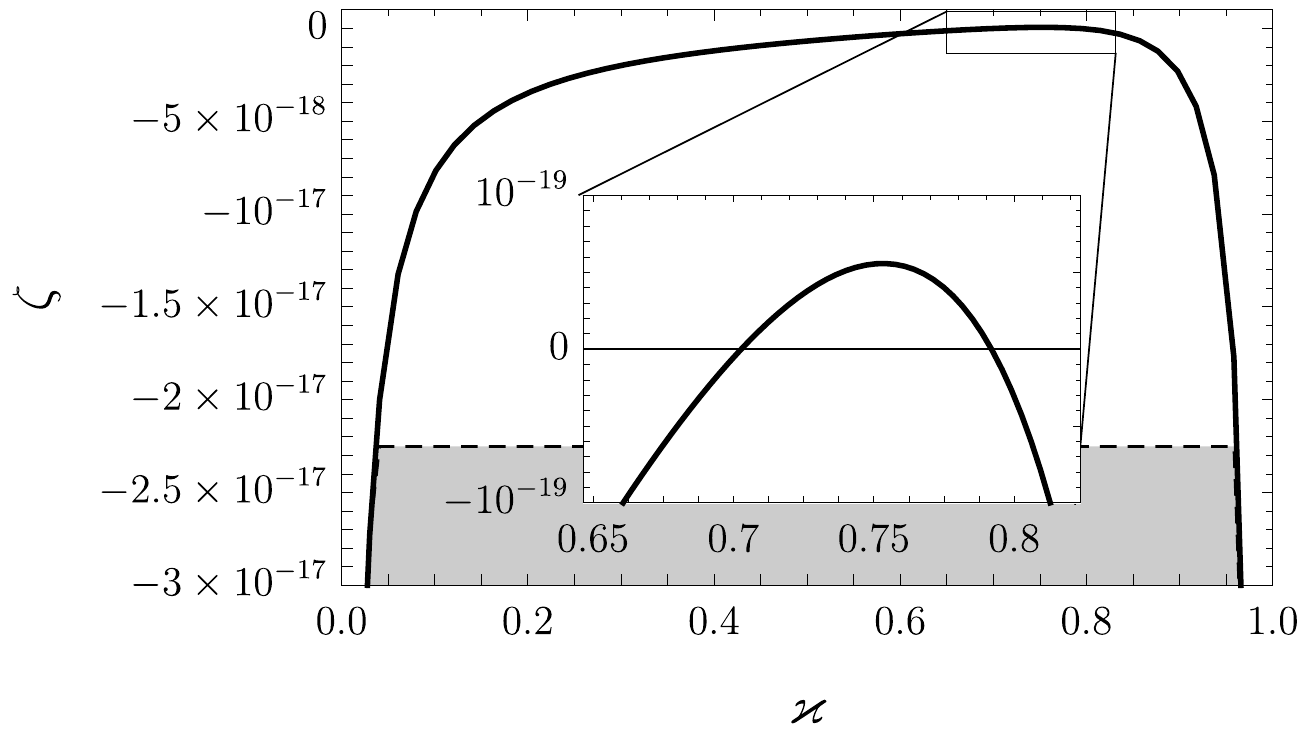}{0.99}
\end{minipage}
\begin{minipage}[t]{0.49\textwidth}
\postscript{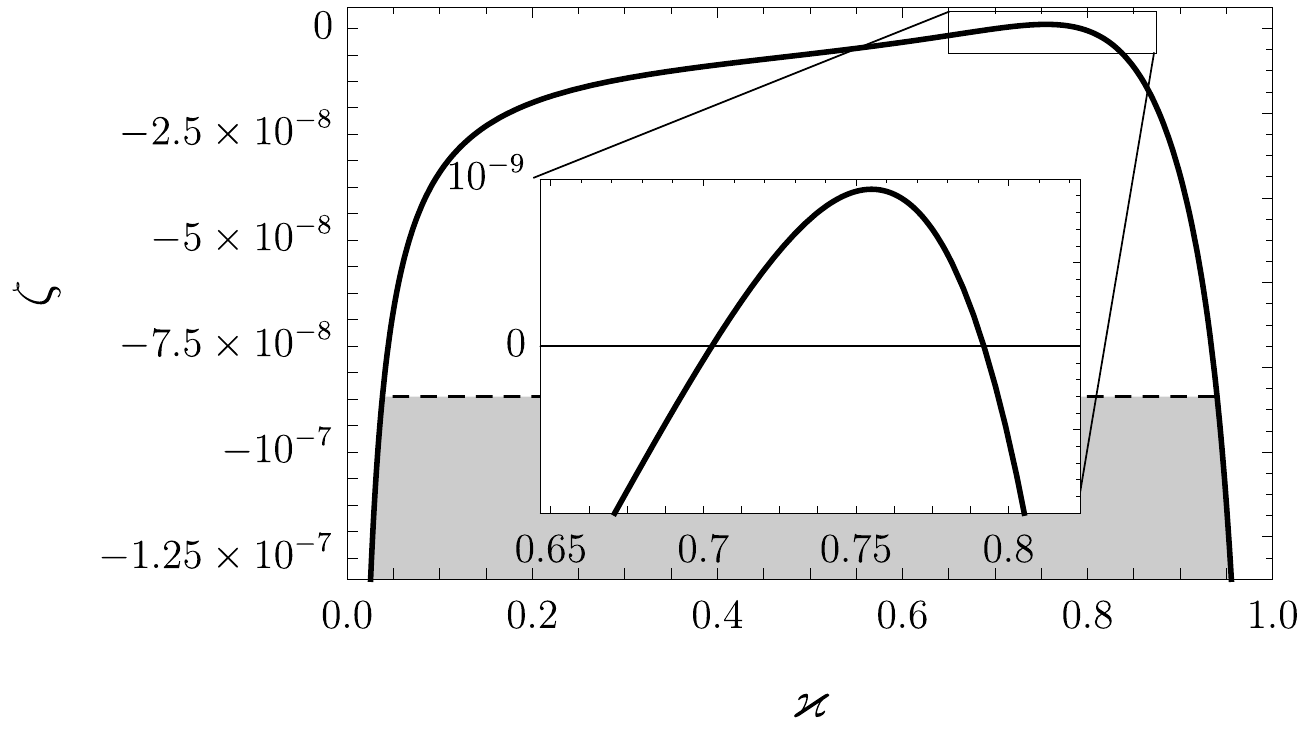}{0.99}
\end{minipage}
\caption{Sensitivity to $\zeta$ as a function of $\varkappa$ for $n=0$
  (left) and $n=1$ (right). We have taken $\varepsilon'_{\rm th} =
  20~{\rm MeV}$ and $E = 10^{9.3}~{\rm GeV}$. The embedded box
  details the restricted interval of $\varkappa$ for which
  $\zeta>0$. The shaded band indicates the region for which $\zeta <
  \zeta_{\rm null}$. \label{fig:3}}
\end{figure*}
We next consider the threshold configuration for a photon with energy
$\varepsilon'_{\rm th} \approx 20\,\mathrm{MeV}$. In the rest frame of
the Sun, the photon energy is $\varepsilon_{\rm th}$ and the UHECR is
boosted with speed $\beta$ in the direction of Earth. For a head on collision, $\mathbf
k$ points in the opposite direction and so the photon energy in the
nucleus rest frame is $\varepsilon' =\gamma(\varepsilon +\beta
k)=\gamma\varepsilon(1+\beta)$. The threshold energy in the rest frame
of the Sun is then \begin{equation}\varepsilon_{\rm
    th}=\sqrt{\frac{1-\beta}{1+\beta}}\varepsilon'_{\rm
    th}.\end{equation} Since $p=\beta E$, we can write
\begin{eqnarray}
\zeta & = &
	\left(\frac
		{[1-(1-\varkappa)A]^2}
		{\varkappa(1-\varkappa)}
		-\frac{4 \beta \,  E\,\varepsilon'_{\rm th}}{m_p^2} \sqrt{\frac{1-\beta}{1+\beta}}
              \right) \nonumber \\
& \times &
              \frac{\left(\beta E/m_p\right)^{-2} \left(\beta
                  E/M_{\rm Pl}\right)^{-n}}{g(\varkappa)} \, .
\label{ventiseis}
\end{eqnarray}
We take $E \approx 10^{9.3}\,{\rm
  GeV}$ and so $\gamma \sim 10^9$. With this in mind, we adopt the
following expansion  
\begin{equation}\sqrt{\frac{1-\beta}{1+\beta}}=\frac{1}{2\gamma}+\mathcal
  O\left(\frac{1}{\gamma}\right)^3,
\label{ventisiete}
\end{equation} and set $\beta \approx 1$ elsewhere. Substituting
(\ref{ventisiete}) into (\ref{ventiseis}) we obtain an expression for
the sensitivity of $\zeta$ as a function of $\varkappa$,
\begin{equation}
\zeta=
\left(\frac
		{[1-(1-\varkappa)A]^2}
		{\varkappa(1-\varkappa)}
		-\frac{2 A \,\varepsilon'_{\rm th}}{m_p}\right) \frac{(m_p/E)^2(M_{\rm Pl}/E)^n}{g(\varkappa)},\label{eq:zeta}
\end{equation} 
where we have used $E=\gamma A m_p$.  As an illustration, in
Fig.~\ref{fig:3} we show the sensitivity for probing
$\zeta$ as a function of $\varkappa$, assuming observation of a few spatiotemporal coincident
  showers near the critical energy.
  
Despite the assumption of Lorentz invariance violation, we want to preserve the time-like character of physical trajectories. For a particle with four momentum $p^\mu$, this means that $p_\mu p^\mu>0$ in a $(+,-,-,-)$ metric signature. Using (\ref{eq:dispersion}) this condition creates a lower bound $\zeta >\zeta_{\rm null}$, with \begin{equation}\zeta_{\rm null}\equiv-A^4\left(\frac{m_p}{E}\right)^2\left(\frac{M_{\rm Pl}}{E}\right)^n,\label{eq:znull}\end{equation} assuming $\beta\approx1$.

The time-like condition is automatically satisfied for positive
$\zeta$. In Fig. \ref{fig:3} we show the limiting value $\zeta_{\rm
  null}$ and the (shaded) prohibited region. We conclude that with a
detection of a few spatiotemporal coincident showers we will be able
to constrain $\zeta$ at the level of $\zeta \sim 5 \times 10^{-20}$ for
$n=0$, and $\zeta \sim 10^{-9}$ for $n=1$.

\begin{figure}\postscript{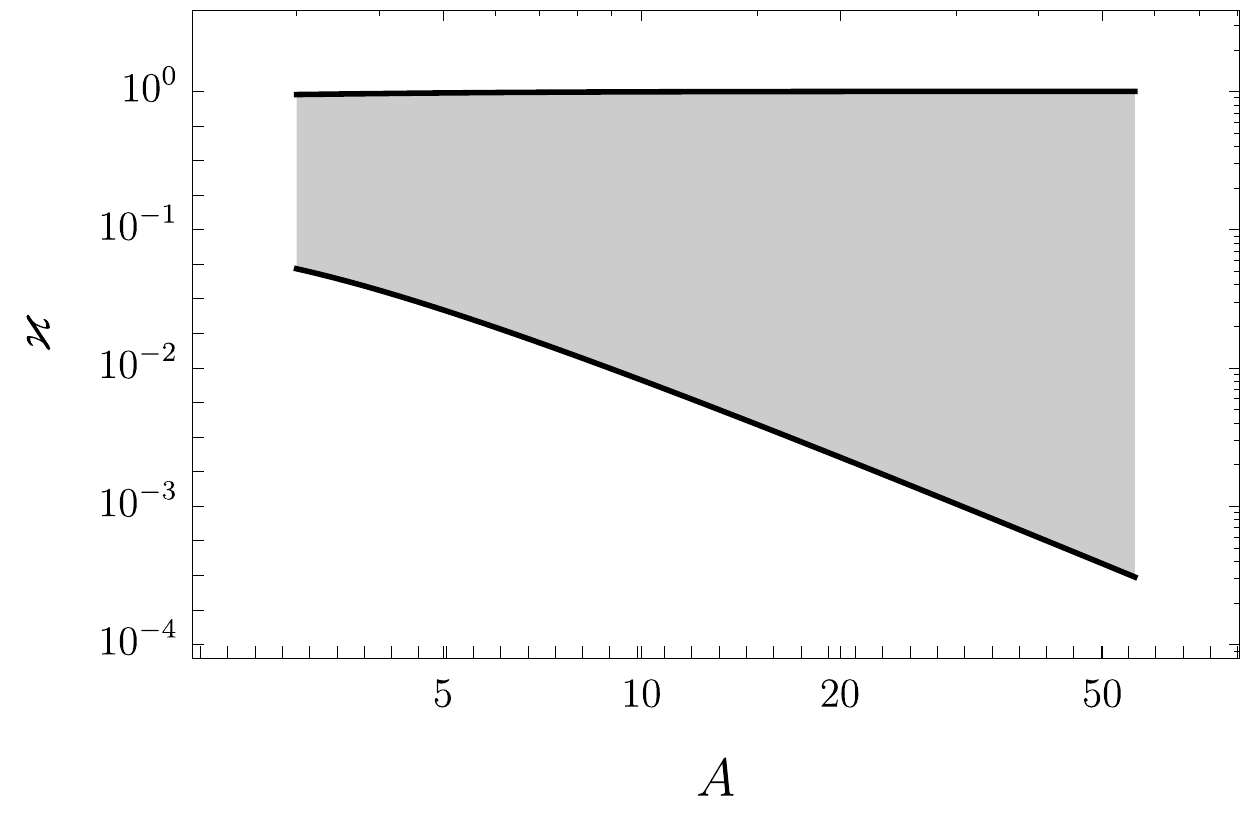}{0.99}\caption{Allowed (shaded) $\varkappa$ region as a function of $A$, for $n=0$. The lower and upper curves are $\varkappa_{\rm min}$ and $\varkappa_{\rm max}$, respectively.}\label{fig:kappalim}\end{figure}

Since $g(\varkappa)\leq0$ for $n=0,1$, using (\ref{eq:zeta}) and
(\ref{eq:znull}) we can rewrite the time-like condition as \begin{equation}
\frac
		{[1-(1-\varkappa)A]^2}
		{2A\varkappa(1-\varkappa)}
		+\frac12A^3g(\varkappa)<\frac{\,\varepsilon'_{\rm
                    th}}{m_p} \, .
\label{eq:timelike_cond}\end{equation}
Using (\ref{eq:timelike_cond}) we study  the dependence on $A$ and
$\varepsilon'_{\rm th}$ of the limiting values $\varkappa_{\rm min}$
and $\varkappa_{\rm max}$, such that the time-like condition is
satisfied for all $\varkappa\in[\varkappa_{\rm min},\varkappa_{\rm
  max}]$. Note that near the limits of the interval $[\varkappa_{\rm min},\varkappa_{\rm max}]$, $d \zeta/ d \varkappa$ is large compared to
$\varepsilon'_{\rm{th}}/m_p$, for $10 \alt \varepsilon'_{\rm th}/{\rm
  MeV} \alt 20$~\cite{Stecker:1998ib}. Thus,  the intervals of $\varkappa$ which
satisfy (\ref{eq:timelike_cond}) barely depend on $\varepsilon'_{\rm
  th}$, which can be assumed to be zero.  For a fixed $n$, the
$\varkappa$ limits only depend on $A$. The values of $\varkappa_{\rm min}$ and $\varkappa_{\rm max}$ for $n=0$ are shown in Fig. \ref{fig:kappalim}. For $n=1$, the values are within a distance of $\sim10^{-2}$ of those for $n=0$. As can be seen, the values are considerably close to $0$ and $1$, with intersections at $[0.04,0.96]$ for $A=4$.

In summary, we have shown that if the photo-disintegration probability
of UHECR nuclei on the solar radiation field is ${\cal O}
(10^{-5.5})$, then the unambiguous observation of the extensive air
showers that would be produced almost simultaneously by the secondary
fragments is within reach of UHECR experiments. This is because our
analysis of spatiotemporal correlations indicates that for angular
scales $\alt 3^\circ$ and a time window of ${\cal O}(10~{\rm s})$ the
signal is background free.  Detection of a few events will be enough
to constrain Lorentz invariant breaking effects in the range $10^9
\alt \gamma \alt 10^{10}$.  Such
a detection will also provide valuable information on the UHECR
nuclear composition, which is independent of the hadronic interaction
models used to describe the development of air showers, and therefore
such information develops complementary to studies of the $X_{\rm
  max}$ distribution and its fluctuations.

\acknowledgments{We would like to acknowledge many useful discussions
  with Tom Weiler, Michael Unger, and our colleagues of the Pierre Auger and POEMMA
  collaborations.  This work has been supported by the U.S. National
  Science Foundation (NSF) Grant No. PHY-1620661 and by the National
  Aeronautics and Space Administration (NASA) Grant No. NNX13AH52G.}



\end{document}